\documentclass[,final]{aipproc}
\layoutstyle{8x11double}

\begin{document}

\title{On resonant and non-resonant origin of double-mode Cepheid pulsation}


\author{R. Smolec}{
  address={Copernicus Astronomical Center, ul.~Bartycka~18, 00-716~Warsaw, Poland}
}

\author{P. Moskalik}{
  address={Copernicus Astronomical Center, ul.~Bartycka~18, 00-716~Warsaw, Poland}
}

\begin{abstract}
Double-mode Cepheid behaviour may arise either from non-resonant or resonant mode interaction. Inclusion of turbulent convection into pulsation codes by Koll\'ath et al. led to stable double-mode F+1O Cepheid pulsation \cite{kollath98}. However, our recent computations indicate, that these models resulted from incorrect neglect of negative buoyancy effects \cite{SiM08b}. Once these effects are taken into account, only some resonant double-mode models could be found. These models do not solve the puzzle of double-mode phenomenon, as they are restricted to narrow parameter ranges. For majority of the observed double-mode Cepheids non-resonant mechanism has to be operational.
\end{abstract}

\maketitle

\section{Introduction}

Most of classical Cepheids are single-mode pulsators, pulsating predominantly either in the fundamental mode (F) or in the first overtone (1O). Few hundreds of double-mode variables are also known. Most of these pulsate simultaneously in two consecutive radial modes, either in fundamental and first overtone (F+1O) or in two lowest order overtones (1O+2O). 

The mechanism responsible for the double-mode behaviour is far from being understood. Necessary condition for double-mode pulsation to occur is simultaneous linear instability of two pulsation modes of interest. We consider fundamental and first overtone modes assuming for simplicity that all other modes are linearly stable. Fundamental mode is linearly unstable (its linear growth rate, $\gamma_0$, is positive) at the red side of the instability strip (IS), while first overtone is linearly unstable ($\gamma_1>0$) at the blue side of the IS. In the middle of the IS lays the potential domain of the double-mode pulsators. As the single-mode Cepheid enters this domain during its evolution, its pulsation state (mode selection) is determined by nonlinear effects. The important quantity to follow is the stability of the single-mode full amplitude (limit cycle) pulsation, with respect to perturbation in the other linearly excited mode. This is expressed through the single-mode stability coefficients, $\gamma_{i,j}$, which describe the stability of the limit cycle pulsation in mode $j$ with respect to perturbation in mode $i$. Positive value of the stability coefficient means that respective single-mode pulsation is unstable and will tend to switch into the other mode. Simultaneous instability of two limit cycles, say, $\gamma_{1,0}>0$ and $\gamma_{0,1}>0$, leads unavoidably to F+1O double-mode pulsation.

There are two mechanism that can lead to simultaneous instability of both limit cycles, non-resonant and resonant \cite{DaK}. In non-resonant mechanism two pulsation modes compete with each other to saturate the pulsation instability. If none of the modes is able to saturate the instability alone, double-mode pulsation arises. In resonant mechanism one of the modes (or both) is resonantly coupled with linearly damped mode.  The damped, parasite mode, can limit the amplitude of the otherwise dominant mode and thus, decrease the stability of its limit cycle. If the mode is destabilized in a parameter range in which the limit cycle of the other linearly excited mode is also unstable, double-mode pulsation arises.

Low order resonances among pulsation modes are rare in a period range occupied by double-mode variables and can affect the mode selection only in a rather narrow parameter range. Therefore, in most of the observed double-mode Cepheids (both F+1O and 1O+2O) non-resonant mechanism should be operational. For many years, nonlinear radiative computations failed to reproduce the double-mode Cepheid behaviour. It was the inclusion of turbulent convection into the model equations that led to success \cite{kollath98}. However, our recent computations \cite{SiM08b} indicate, that the computed non-resonant double-mode models resulted from incorrect physical assumption adopted by Koll\'ath et al. \cite{kollath98}. We discuss these results below, first, providing a brief overview of turbulent convection recipes used in pulsation hydrocodes.

\section{Turbulent Convection Recipes}

There are several models of turbulent convection and its interaction with pulsation. However, only the simplest models are suitable for nonlinear computations. In these models, generation of turbulent energy, $e_t$, is described by one additional equation, which has to be solved together with momentum and energy equations. In the Kuhfu\ss{} model \cite{kuhfuss}, which we will discuss here, these equations are following (see \cite{SiM08a}):
\begin{equation}\frac{{\rm d}u}{{\rm d}t}=-4\pi r^2\frac{\partial}{\partial M_r}(p+p_t)+U_q-\frac{GM_r}{r^2},\end{equation}
\begin{equation} \frac{{\rm d}E}{{\rm d}t}+p\frac{{\rm d}V}{{\rm d}t}=-\frac{\partial(L_{\rm r}+L_{\rm c})}{\partial M_r}-C,\end{equation}
\begin{equation} \frac{{\rm d}e_t}{{\rm d}t}+p_t\frac{{\rm d}V}{{\rm d}t}=-\frac{\partial L_{\rm t}}{\partial M_r}+E_q+C.\end{equation}
Velocity $u$ is time derivative of radius, $u={\rm d}r/{\rm d}t$, $M_r$ is mass enclosed in radius $r$, $V$ is specific volume, $p$ is gas pressure, $p_t$ is turbulent pressure, $L_{\rm r}$, $L_{\rm c}$ and $L_{\rm t}$ are radiative, convective and turbulent luminosities respectively. $U_q$ and $E_q$ are viscous momentum and energy transfer rates, respectively, and $C$ is the term coupling the turbulent and internal energy equations. Its form is following:
\begin{equation}C=S-D-D_{\rm r}.\end{equation}
$S$ is the turbulent source function, $D$ is turbulent dissipation function and $D_{\rm r}$ is radiative cooling function. The model is a simple phenomenological recipe, containing 8 order of unity free parameters. Nevertheless, all the terms have clear physical interpretation. Turbulent source function represents the buoyant forces acting on the convective eddy, responsible for its acceleration. It is proportional to superadiabatic gradient, $S\sim Y$, $Y=\nabla-\nabla_{\rm a}$. According to Schwarzschild criterion for convective instability, in stellar layers in which $Y>0$, buoyancy represented in the source term drives the convective motions. In case of convective stability, $Y<0$, turbulent source function is negative and contributes to the damping of turbulent energies. Turbulent dissipation term models the turbulent cascade -- the decay of turbulent eddies into smaller and smaller scales down to molecular scale where energy is dissipated. Radiative cooling terms damps the turbulent energies also, as it represents the radiative losses during the motion of the eddy. Viscous momentum and energy transfer rates describe the coupling between turbulent motions and mean, pulsation induced gas motion. Turbulent energy is always generated at the cost of pulsation energy, thus, eddy-viscous damping is an important factor limiting the pulsation amplitude. For more detailed description of the model, we refer the reader to \cite{SiM08a}.

Described model was adopted in Warsaw pulsation codes \cite{SiM08a}. Following notation introduced by \cite{SiM08a} we denote this model, default in all our computations, by NN. Very similar model was adopted in Florida-Budapest hydrocode (eg. \cite{kollath98}, \cite{kollath02}), which was used to compute the first robust double-mode Cepheid models. In the model adopted in that code, superadiabatic gradient, $Y$, was restricted to positive values. Consequently turbulent source function, $S\sim Y_+$, was set equal to zero in convectively stable regions. Such assumption accounts to the neglect of a very important physical effect acting in convectively stable regions, namely to the neglect of negative buoyancy. In our opinion such neglect is not justified. Buoyant forces do not vanish in convectively stable regions. In fact, it is the negative buoyancy that restores the convective stability. Discussed assumption leads to different convection model (PP in the following), which we implemented in our codes as an option. Detailed comparison of models computed with NN and PP convection revealed, that the double-mode behaviour observed in PP models resulted entirely from unphysical neglect of negative buoyancy effects.

\section{Non-resonant double-mode Cepheid models -- Revisited}

\begin{figure}
  \includegraphics[height=.23\textheight]{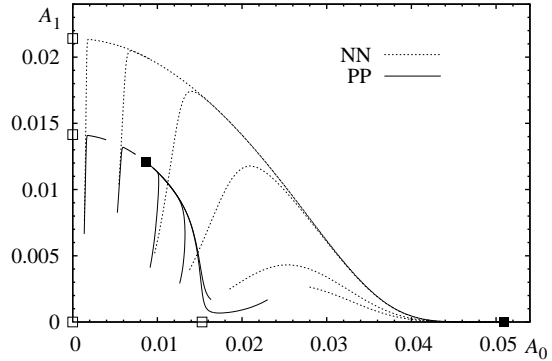}
  \caption{Hydrodynamic integrations for particular Cepheid model with different initial conditions.\label{traj}}
\end{figure}

To search and analyze the mode selection we adopted the methods described by \cite{kollath02} (see Fig.~\ref{traj}). For several models along a sequence of eg. constant luminosity and varying temperature, direct nonlinear integrations were performed. Each model was kicked with several initial conditions and time evolution of mode amplitudes was followed through the analytical signal method. Resulting trajectories for particular model adopting NN convection are shown with dashed lines in Fig.~\ref{traj}. Model trajectories were fitted with fifth order amplitude equations, which allowed to find all the possible pulsation states and their stability. Stable pulsation states (attractors) are marked with solid squares in Fig.~\ref{traj}, while unstable with open squares. Using interpolation, modal selection may be found at each temperature along a sequence. For detailed description of the method, we refer the reader to \cite{SiM08b}.

In Fig.~\ref{traj} two sets of trajectories are plotted. Dashed lines correspond to model computed with NN convection, while solid lines, to model of exactly the same parameters, but computed with PP convection. Most striking difference concerns the mode selection. For PP model, stable double-mode solution exists, while for NN model of the same parameters only fundamental mode limit cycle is stable. The other difference is a significant reduction of mode amplitudes of PP model, as compared to the NN model. Particularly, amplitude of the fundamental mode is reduced very strong. Described differences are typical. They do not depend on exact parameters of the model, either convective or physical. Particularly, along model sequences computed using PP convection, stable double-mode solutions are easily found, while no such solution was found for models computed using NN convection, despite extensive search \cite{SiM08b}. As analyzed by \cite{SiM08b} and briefly summarized below, the differences between NN and PP models result from the neglect of negative buoyancy effects in PP model. As such neglect is not justified, stable double-mode behaviour observed in PP models is not correct.

\begin{figure}
  \includegraphics[height=.44\textheight]{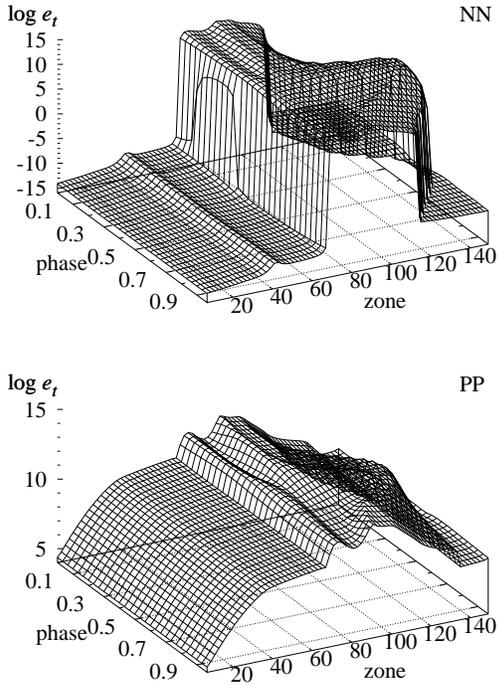}
  \caption{Profiles of turbulent energy during one cycle of full amplitude fundamental mode pulsation.\label{profiles}}
\end{figure}

\begin{figure}
  \includegraphics[height=.33\textheight]{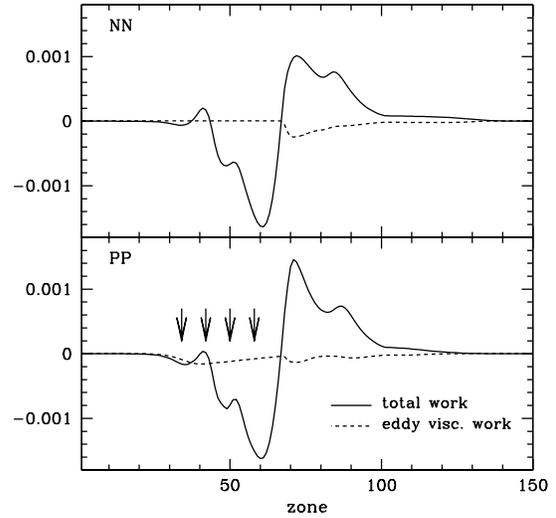}
  \caption{Nonlinear work integrals for full amplitude fundamental mode pulsation.\label{work}}
\end{figure}

We focus our attention on the simplest convective model, that is we neglect the turbulent flux, turbulent pressure and radiative losses. We stress however, that inclusion of these effects does not change the qualitative conclusions presented below. With these simplifications, turbulent energy equation is following:
\begin{equation} \frac{{\rm d}e_t}{{\rm d}t}=S-D+E_q\end{equation}
The crucial difference between NN and PP models appears in convectively stable regions, where $Y<0$. In these regions of NN model, turbulent source function is negative and damps the turbulent energies. In PP model it is set equal to zero. In both models, turbulent dissipation term, $D$, always damps the turbulent energies, while eddy-viscous term, $E_q$, always drives the turbulent energies. Consequences of different treatment of turbulent source function are presented in Figs.~\ref{profiles} and \ref{work}.

In Fig.~\ref{profiles} we plot the profiles of turbulent energy during one cycle of fundamental mode pulsation. Turbulent energies are plotted in logarithmic scale to highlight the differences in convectively stable regions, barely visible in a linear scale. Convectively stable, internal layers extend below zone 70 in both NN and PP models. Here the differences in turbulent energy profiles are visible. In NN model, turbulent energies fall rapidly as superadiabatic gradient changes its sign. Turbulent motions are braked effectively by negative buoyancy represented in the source term. In internal layers of NN model, turbulent energies are negligible (but nonzero, see \cite{SiM08b}). In case of PP model situation is different. As superadiabatic gradient changes its sign, turbulent energies fall, but only by three orders of magnitude, as compared to the values in the center of the convection zone. Then, at the absence of negative source function, the balance between the turbulent dissipation and eddy-viscous forces sets the turbulent energies at a level of $10^8-10^{10}$\thinspace{}erg/g, in a large region, extending to more than 6 local pressure scale heights below the envelope convective zone (zones 40--70). These turbulent energies are generated at the cost of pulsation through the eddy-viscous forces. Although these energies seem small, their effect on pulsation dynamics is significant, which is best visible in work integrals displayed in Fig.~\ref{work}. In the internal layers of the PP model, strong eddy-viscous damping is present (arrows in Fig.~\ref{work}). No such damping is visible in convectively stable layers of NN models, as turbulent energies are not significant there. As a result amplitudes of pulsation modes computed with PP convection are much smaller, which is clearly visible in Fig.~\ref{traj}. Amplitude of the fundamental mode is always reduced stronger, than the amplitude of the first overtone due to the different properties of these modes in the interior of the model\footnote{Below the pulsation node in the velocity profile of the first overtone mode, eddy-viscous driving is very small, as $E_q$ term is proportional to the square of spatial derivative of velocity. Consequently, internal eddy-viscous damping in convectively stable regions of the PP model is weaker for the first overtone mode (and overtones in general).}.

\begin{figure}
  \includegraphics[height=.22\textheight]{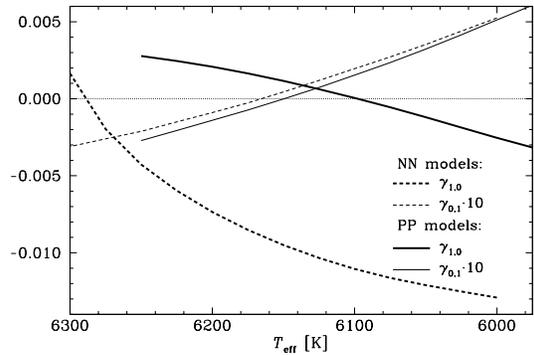}
  \caption{The run of stability coefficients along a sequence of Cepheid models of constant luminosity.\label{stab}}
\end{figure}

Strong reduction of fundamental mode amplitude has significant effect on its limit cycle stability (described by $\gamma_{1,0}$). Lower the amplitude of the mode, more unstable it is against perturbation in the other mode. The run of stability coefficients of the fundamental and first overtone limit cycles, along a sequence of models of constant luminosity is shown if Fig.~\ref{stab}. The qualitative scenario for both NN and PP models is similar and consistent with what we expect from non-resonant mode interaction. First overtone becomes more unstable ($\gamma_{0,1}$ increases) toward the red, while fundamental mode becomes more unstable ($\gamma_{1,0}$ increases) toward the blue. For NN model sequence fundamental mode limit cycle becomes stable at high temperatures, at which first overtone limit cycle is still firmly stable. No stable-double mode solution is possible. Instead, we observe a mode selection typical for radiative computations: domains of single-mode first overtone pulsation (blue side of the IS) and fundamental mode pulsation (red side) are separated by either-or domain in which pulsation in either mode is possible, depending on initial conditions (direction of evolution). Situation is very different for PP model sequence. Due to its strongly reduced amplitude, fundamental mode is more unstable in PP sequence. It remains unstable at relatively low temperatures, at which first overtone is already unstable. Consequently in a particular temperature range both limit cycles are unstable, giving rise to stable double-mode pulsation. However, the cause of the observed double-mode behaviour is unphysical. It arose due to unphysical neglect of negative buoyancy effects.

\section{Resonant double-mode Cepheid models}

Despite rather disappointing results presented in the preceding Section, we have found several stable double-mode models with NN convection, including negative buoyancy effects. In these models resonant mechanism is operational. Computed models include both F+1O Cepheids, with metallicities corresponding to Galaxy, and 1O+2O Cepheids found during the computations intended to model the double-overtone Cepheids in the LMC. In the former case (F+1O), the three-mode resonance of the type $2\omega_1=\omega_0+\omega_2$ is the cause of the double-mode pulsation, while in the latter case (1O+2O), the 2:1 resonance, $2\omega_1=\omega_5$ is operational. Details of the models will be published in the nearest future (Smolec \& Moskalik, in prep.), here we present the preliminary results.

\begin{figure}
  \includegraphics[height=.22\textheight]{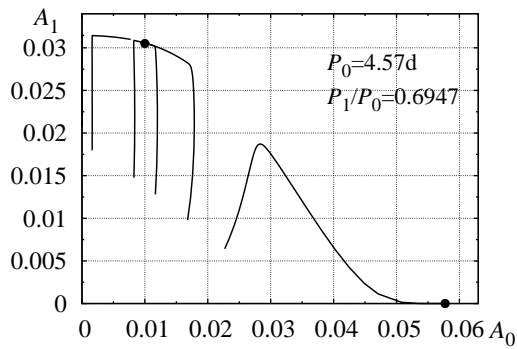}
  \caption{Hydrodynamic integrations for particular Cepheid model crossing the $2\omega_1=\omega_0+\omega_2$ resonance center.\label{parrezo}}
\end{figure}

In Fig.~\ref{parrezo} we display the results of numerical integration of one Cepheid model lying close to the three-mode resonance center. Stable double-mode attractor is clearly visible. It coexists with stable fundamental mode attractor and this mode selection is a rule in all double-mode models of this type that we found. The model is very close to the resonance center, the proximity parameter, $\Delta=2\omega_1/(\omega_0+\omega_2)=0.9995$. Period and period ratios for this model are given in Fig.~\ref{parrezo} and they are also typical. Somewhat too low period ratio, as compared with observation, can be compensated by slight decrease of model metallicity. In all our model sequences double-mode behaviour was restricted to a narrow temperature range (at most 50\thinspace{}K, usually below 20\thinspace{}K). Computed double-mode models always fulfilled two conditions: (\emph{i}) they were always located very close to the resonance center and (\emph{ii}) in between the single-mode fundamental mode pulsation domain (to the red) and F/1O either-or domain (to the blue), which is rather typical for non-resonant models. At the moment, we do not have explanation why (\emph{ii}) is necessary for the double-mode behaviour to occur.

\begin{figure}
  \includegraphics[height=.22\textheight]{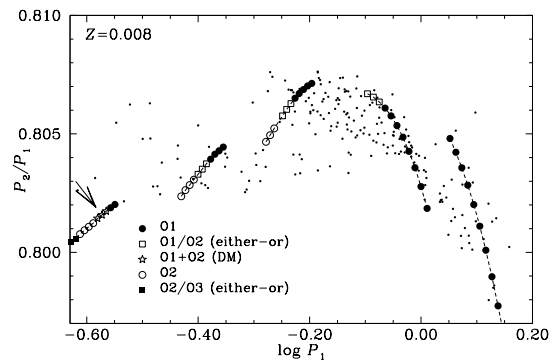}
  \caption{Petersen diagram for the double-overtone LMC Cepheids. Different symbols indicate the mode selection inferred from nonlinear computations.\label{lmc}}
\end{figure}

Other example of the double-mode behaviour was found during the model survey intended to reproduce the observed double-overtone Cepheids in the LMC. Our results are summarized in Fig.~\ref{lmc}, showing the Petersen diagram for the double-overtone LMC Cepheids \cite{Sosz08}. Mode selection was computed along evolutionary tracks of 2.5${\rm M_S}$, 3.0${\rm M_S}$ and 3.5${\rm M_S}$, corresponding to the first crossing of the IS (three leftmost tracks in Fig.~\ref{lmc}) and along two horizontal paths for models with 3.0${\rm M_S}$ and 3.5${\rm M_S}$ and luminosity increased by $\Delta\log L=0.4$ relative to evolutionary track (remaining two tracks in Fig.~\ref{lmc}). Such luminosity increase was intended to model the core helium burning phase (blue loop). For detailed results of linear modeling and challenges it represent for stellar evolution theory, we refer the reader to \cite{DaS}, here focusing on mode selection only.

It is clearly visible that inferred modal selection disagree with observation. Double-mode models were found only at the shortest periods.  Linear analysis indicate, that the origin of the double-mode pulsation is most likely connected with the 2:1 resonance between the excited first overtone and linearly damped fifth overtone. Indeed, resonance center, indicated with arrow in Fig.~\ref{lmc} lays in the middle of the double-mode domain. Resonant mechanism can be operational only at shorter periods. At longer periods ($\log P_1>-0.1$) no low order resonances between pulsation modes are found and a non-resonant mechanism has to be operational. As in case of F+1O variables, no non-resonant 1O+2O models were found.
 
\section{Conclusions}

Presented results are rather disappointing and significantly change the picture emerging from the recent works of the Florida-Budapest group (eg. \cite{kollath98}, \cite{kollath02}, Buchler, these proceedings). We have shown that their non-resonant F+1O double-mode Cepheid models resulted from the incorrect neglect of negative buoyancy effects. When these effects were included in model equations, no non-resonant double-mode behaviour was found. Although we have found some resonant double-mode models including negative buoyancy in model computations, these models do not offer the solution to the problem of modelling the double-mode Cepheids. For majority of the stars a non-resonant mechanism has to be operational. It seems that the better treatment of convection/pulsation coupling is necessary to solve the puzzle of double-mode Cepheids.

\begin{theacknowledgments}
This work has been supported by the Polish MNiSW grant No 1 P03D 011 30.
\end{theacknowledgments}

\bibliographystyle{aipproc}

\begin{thebibliography}{9}

\bibitem{kollath98}
Z.~Koll\'ath, et al., \emph{ApJ}, \textbf{502}, 55--58 (1998).

\bibitem{SiM08b}
R.~Smolec and P.~Moskalik, \emph{Acta Astronomica}, \textbf{58}, 233--261
(2008).

\bibitem{DaK}
W.~A.~Dziembowski and G.~Kov\'acs, \emph{MNRAS}, \textbf{206}, 497--519 (1984).

\bibitem{kuhfuss}
R.~Kuhfu\ss{}, \emph{A\&A}, \textbf{160}, 116--120 (1986).

\bibitem{SiM08a}
R.~Smolec and P.~Moskalik, \emph{Acta Astronomica}, \textbf{58}, 193--232
(2008).

\bibitem{kollath02}
Z.~Koll\'ath, et al., \emph{A\&A}, \textbf{385}, 932--939 (2002).

\bibitem{Sosz08}
I.~Soszy\'nski et al.,  \emph{Acta Astronomica}, \textbf{58}, 163--185 (2008).

\bibitem{DaS}
W.~A. Dziembowski and R.~Smolec, \emph{Acta Astronomica}, \textbf{59},
 19--31 (2009).

\end{thebibliography}

\end{document}